\documentclass[a4paper]{article}
\usepackage[utf8]{inputenc}
\usepackage{color}
\usepackage{amsmath}
\usepackage{amssymb}
\usepackage{amsfonts}
\usepackage{amsthm}
\usepackage{graphicx}
\usepackage{nicefrac}
\usepackage{enumerate}
\usepackage{natbib}

\usepackage{geometry} 
\newtheorem{example}{Example}
\newtheorem{assumption}{Assumption}
\newtheorem{application}{Application}
\theoremstyle{remark}
\newtheorem{remark}{Remark}
\usepackage{url}

\newcommand{\datasetspace}{\mathcal{D}}
\newcommand{\objectspace}{\mathcal{X}}
\newcommand{\generator}{\mathcal{G}}

\newcommand{\synthdata}{{D^{(s)}}}
\newcommand{\realdata}{{D^{(r)}}}
\newcommand{\linspan}[1]{\mathop{\mathrm{span}}\big\langle#1\big\rangle}

\title{A Framework for Auditable Synthetic Data Generation}

\author{
Florimond Houssiau$^{1,2}$, Samuel N. Cohen$^{1,3}$, Lukasz Szpruch$^{1,4}$, Owen Daniel$^{2}$,\\ Michaela G. Lawrence$^{2}$, Robin Mitra$^{5}$, Henry Wilde$^{2}$, Callum Mole$^{1,2}$\vspace{.1cm}\\
\small $^{1}$The Alan Turing Institute, $^{2}$The Office for National Statistics, $^{3}$University of Oxford,\\ \small $^{4}$University of Edinburgh, $^{5}$University College London.
}

\begin{document}

\maketitle

\begin{abstract}
Synthetic data has gained significant momentum thanks to sophisticated machine learning tools that enable the synthesis of high-dimensional datasets. However, many generation techniques do not give the data controller control over what statistical patterns are captured, leading to concerns over privacy protection. While synthetic records are not linked to a particular real-world individual, they can reveal information about users indirectly which may be unacceptable for data owners.
There is thus a need to empirically verify the privacy of synthetic data --- a particularly challenging task in high-dimensional data.
In this paper we present a general framework for synthetic data generation that gives data controllers full control over which statistical properties the synthetic data ought to preserve, what exact information loss is acceptable, and how to quantify it.
The benefits of the approach are that (1) one can generate synthetic data that results in high utility for a given task, while (2) empirically validating that only statistics considered safe by the data curator are used to generate the data. We thus show the potential for synthetic data to be an effective means of releasing confidential data safely, while retaining useful information for analysts.
\footnote{Our code is available at \url{https://github.com/alan-turing-institute/sdg-auditing}.}

\end{abstract}

%% Content
\section{Introduction}

The ability to access accurate, detailed data in real time is vital to well-informed decision-making. As the appetite for large data sources increases, so do concerns about the ethical and responsible use of such data.
Failure to protect confidentiality can cause significant harm to individuals or groups impacted by disclosed information, and regulatory frameworks (such as the General Data Protection Regulation in the EU) are increasingly being introduced to mitigate this risk.
Statistical disclosure control methods address this problem by altering values in the data to mask their true values and thus limit disclosure leaks.
However, ensuring that the released data protects user privacy while maintaining its utility is challenging, especially for high-dimensional data~(\cite{ohm2009broken}).

 \cite{rubin1993statistical} proposed using synthetically generated data as a tool to enable data analysis while mitigating disclosure risk.
Synthetic data has gained significant momentum thanks to sophisticated machine learning tools that enable the synthesis of high-dimensional datasets.
However, while synthetic records are not linked to real-world individuals, a synthetic dataset can still reveal sensitive information about users~(\cite{stadler2021synthetic}).

Synthetic data generators (SDGs) are typically designed to balance three properties: utility, fidelity, and privacy~(\cite{jordon2022synthetic}).
Utility refers to the accuracy of specific inferences from the synthetic data, while fidelity evaluates ``how close'' the distribution of synthetic records is to that of real records.
Privacy measures the risk of disclosure due to the release of sensitive data, and is typically measured through formal definitions such as Differential Privacy~(\cite{dwork2006calibrating}) or adversarial approaches~(\cite{stadler2021synthetic,houssiau2022tapas}).
Prior work has mostly focused on the \textit{fidelity--privacy} tradeoff, assuming that high fidelity entails high utility.

We argue that synthetic datasets should be designed with a focus on utility.
For this, the information extracted from the real data to construct synthetic data should be chosen transparently, with a specific task in mind.
Furthermore, we introduce the additional concept of {\em auditability}: the data holder needs to be able to verify that only statistics it is willing to release are revealed by the SDG, and that no additional information leakage is occurring.
Auditability requires (1) that synthetic data comes with a ``generator card'' that transparently states what information was (and was not) used to generate the data, and (2) data holders can verify that the generator card is correct.
A sound framework for auditing SDGs allows data-holders to release synthetic data with confidence.

In this paper we conceptualise a process encompassing the whole pipeline required to permit release of outputs from confidential data via a synthetic data generator.
We build on prior work to propose a framework where (1) analysts explicitly select the statistics used to construct the synthetic data, and (2) data holders can verify that the synthetic data reveals no further information from the real data.
We introduce the concept of a \textit{generator card}, a transparent statement that specifies which statistics a generator is permitted to use when producing synthetic data.
We then propose a mathematical framework to audit the generator for any unintended information leaks. Bridging theoretical guarantees of privacy with practical applications is crucial, as releasing synthetic data carries the risk of privacy violation, and is the main barrier to the wide-scale adoption of this technology.

\section{Framework}
\label{sec:framework}

We argue for a utility-first approach, where synthetic data is designed with a specific task (or set of tasks) in mind. This allows us to determine the required statistical fidelity and privacy constraints needed for trustworthy synthetic data generation.
To illustrate this idea, consider the following scenario: an \textit{analyst} (A) comes to the \textit{data controller} (B) to perform analysis on B's private data. A cannot access the data directly, but B offers to run A's code (once) on the data. In order to design their code and verify the approach taken is of interest, A requests synthetic data that replicates a set of statistics of the real data. A can then select the right statistical tool for the task at hand and request B to run the code on real data to validate the analysis. 

Prior work by~\cite{mckenna2021winning} proposed the Select--Measure--Generate framework, where synthetic data is produced by (1) selecting statistics of the real data to preserve, (2) measuring these statistics (with added noise), and (3) finding a distribution with these statistics, from which synthetic records are sampled.
We build on this framework by explicitly requiring the Select step to be performed before seeing the data, by agreement between the data controller and analysts, and adding an explicit Auditing step.
We summarise our framework in three key steps.

\begin{itemize}
    \item \textbf{Select.} The analyst and data controller agree on a set of statistics $\Phi$ that (1) should be able to approximately solve a specific set of tasks, and (2) are deemed safe to release by the data controller. We call these ``safe'' statistics.
    \item \textbf{Generate.} A generator $\generator$ is designed (by either party, or a third-party) that produces synthetic data $\synthdata$ from a private, real dataset $\realdata$. This generator should rely exclusively on the safe statistics, and can include some randomisation/imprecision.
    \item \textbf{Audit.} The data controller verifies that the generator does not use any non-safe statistical information about the real data. The data controller is here assumed to have black-box access to the generator.
\end{itemize}

This approach enables the analyst to obtain high utility synthetic data, while allowing the data controller to avoid excessive privacy risks.
Enabling a data holder to evaluate an SDG before granting data access is critical for this technology to be used safely.
Furthermore, in this framework, one can attach a ``label'' to a synthetic dataset stating which statistics went into its construction, as well as additional metadata.
We call such label a {\em generator card}, which conveys the necessary information to verify the synthesis process was fit for purpose.
This idea is inspired by \textit{model cards} developed for machine learning models by \cite{mitchell2019model}.
Generator cards make synthetic data generators more transparent to stakeholders unfamiliar with synthetic data generation.

Another practical advantage of this framework is that it recognises that many public sector organisations (e.g. National Statistics Offices) already release summary statistics from sensitive data. In this setting, synthetic data could be generated using only information which an organisation is happy to put in the public domain. 

\section{Auditing Generators}
In this section, we propose an \textit{auditing procedure} that verifies whether a generator only uses ``safe'' statistics in $\Phi$, and its connection with a generator card.
This allows data controllers to verify that a synthetic dataset reveals only information they are comfortable with.

\subsection{Formal Framework and Generator Cards}

We now present a formal overview of the problem of identifying information leakage in a synthetic data generator. Our construction aims to infer whether a specific generator $\generator$ relies exclusively on a set of ``safe'' statistics to generate data.

Denote by $\objectspace$ the set of all possible records, i.e.~the space in which our data lives, and by $\datasetspace= \cup_{n\in \mathbb{N}} \objectspace^n$ the set of all datasets composed of records in $\objectspace$.
We study an abstract setting where a synthetic data generator $\generator$ is a map from (real) datasets to distributions over (synthetic) datasets, $\generator: \datasetspace \rightarrow \mathcal{P}(\datasetspace)$ (where $\mathcal{P}(X)$ denotes the set of probability distributions over $X$).
That is, for any real dataset $d = \realdata$, the generator defines a rule for randomly generating new datasets $\synthdata \sim \generator(d)$, which should agree with the real data in some (but not all) respects. A simple example (which we consider in more detail below) is where we learn a distribution over possible records $P\in \mathcal{P}(\objectspace)$, and $\generator(d)$ is the distribution over datasets corresponding to taking an i.i.d.~sample of a given size, where each random observation is drawn using $P$.

In what follows, we will represent datasets using an exhaustive set of statistics.
Note that we can study a dataset $\{x_1,\ldots,x_n\} \in \datasetspace$ via its empirical distribution $\frac{1}{n}\sum_{i=1}^n \delta_{x_i} \in  \mathcal{P}(\objectspace)$; hence a dataset can be seen as an empirical distribution over data entries, that is, $\datasetspace \equiv \mathcal{P}(\objectspace)$, up to rescaling by sample size.   
In order to work with the space $\mathcal{P}(\objectspace)$ more easily, we parameterize it using a set of statistics $\Theta:\mathcal{P}(\objectspace)\to S_\Theta \subset \mathbb{R}^m$. We will assume this statistical parameterization is invertible, that is, for any $\theta\in S_{\Theta}$, there is a unique corresponding distribution over $\objectspace$.

\begin{example} \label{ex1}
Suppose we have data on three binary characteristics of individuals (denoted $A,B,C$). An individual's data $d=(d_A, d_B, d_C)$ is then a point in $\{0,1\}^3 = \objectspace$. If our dataset is $D$, the empirical distribution is given by
\[\hat{\mathbb{P}}(a, b, c) = \#\{d\in D: d_A=a, d_B=b, d_C=c\}\Big/|D|.\]
and $\hat{\mathbb{P}}$ is in $\mathcal{P}(\objectspace)$. We can parameterize $\mathcal{P}(\objectspace)$ by a set of statistics $\theta$ in $\mathbb{R}^7$, for example 
\[\theta = \big\{\hat{\mathbb{P}}(a,b,c)|a,b,c \in \{0,1\}\text{ with }a+b+c<3\big\}.\]
The set of valid values for $\theta$ is a subset of the unit cube $\mathbb{R}^7$ (given restrictions such as nonnegativity of probabilities% and $\hat{\mathbb{P}}(a,b) < \min \{\hat{\mathbb{P}}(a), \hat{\mathbb{P}}(b)\}$ must be observed
), which we denote $S_\Theta$.
\end{example}

Suppose that, within our exhaustive set of statistics, there are some statistics which are deemed ``safe'' to release by the data controller.
For simplicity, we assume that $\Phi\equiv \mathbb{R}^{m_s}$ is a linear space, that is, we can reveal a set of statistics linear in $\Theta$.
Formally, this corresponds to a decomposition of $\mathbb{R}^m$ into $\Phi\times \Phi^\perp$, where $\Phi$ corresponds to safe statistics, and $\Phi^\perp$ corresponds to statistics which must be kept private.
Importantly, such linear statistics could include histograms and $k$-way marginal distributions, 
and thus cover a large range of meaningful tasks\footnote{A useful example is when certain observations are impossible. The statistic `these impossible observations do not appear in our dataset' can then be included in the ``safe'' statistics, rather than requiring us to find a parameterization which excludes them by construction.}
We write $\phi:\datasetspace \to \Phi$ for the map from a dataset to its safe statistics, corresponding to $\phi = \pi_\Phi\circ \Theta$, where $\pi_\Phi$ is the projection map in $\Phi\times\Phi^\perp$.

\begin{example}[Example \ref{ex1} cont.]\label{ex2}
In the same setting as above, suppose we are happy to reveal the proportion of our sample with each characteristic, but not the joint distribution. Our $\Phi$-statistics are $\{\hat{\mathbb{P}}(a), \hat{\mathbb{P}}(b), \hat{\mathbb{P}}(c)\}\in \mathbb{R}^3$. This corresponds to a 3-dimensional subspace of $\mathbb{R}^7\supset S_\Theta$; we denote the remaining dimensions of the space as $\Phi^\perp$. The map $\phi$ takes a dataset (or empirical distribution) and gives the single-category probabilities in $\Phi$.
\end{example}

\begin{example}\label{ex3}
Suppose we have data on a characteristic taking values in $\objectspace=\{0,1,...,10\}$. The data controller is willing to reveal the data's mean and variance, but not higher moments. We describe $\mathbb{P}\in\mathcal{P}(\objectspace)$ by its moments $\Theta(\mathbb{P})=(E_\mathbb{P}[X^k])_{k\leq 10}$, which we decompose into $\Phi = \{(E_\mathbb{P}[X], E_\mathbb{P}[X^2])\}_{\mathbb{P}\in\mathcal{P}(\objectspace)}$ and $\Phi^\perp = \{(E_\mathbb{P}[X^k])_{3\leq k<10} \}_{\mathbb{P}\in\mathcal{P}(\objectspace)}$.
\end{example}
From the perspective of the data controller,  the generator $\generator:\datasetspace \to \mathcal{P}(\datasetspace)$ should depend only on the statistics $\Phi$, and be completely independent of statistics  $\Phi^\perp$; this guarantees the data generated can only reveal safe statistics. We call such a generator $\generator$ \emph{decomposable}: its output is the same for all inputs with the same $\Phi$-statistics as the real data:
\[\generator(d) = \generator(\realdata) \text{ for all }d\text{ such that }\phi(d) = \phi(\realdata).\big.\]
%This implies that $\generator$ can be decomposed as $\generator_\phi\circ \phi$, for some $\generator_\phi:\Phi \to \mathcal{P}(\datasetspace)$.
We define the \textit{generator card} as the triplet $\mathcal{C} = \langle \psi,\Phi,\generator\rangle$, where $\psi = \phi(\realdata)$, that describes the information needed to verify whether a generator is decomposable.

\subsection{Auditing a generator in theory}

Given a generator card $\mathcal{C} = \langle \psi,\Phi,\generator\rangle$, where 
\[    \psi = \text{safe statistics of real data},\quad
    \Phi = \text{space of safe statistics},\quad
    \generator = \text{data generator},
\]
we wish to determine whether $\generator$ is decomposable. We call this process \textit{auditing the generator (card)}, as it aims to verify whether the card is a truthful description of the generator. The challenge is that we usually will only have access to samples from $\mathcal{G}(d)$, but can test using a variety of training datasets $d$, not only the real dataset $\realdata$. We assume that $\psi=\phi(\realdata)$ is a truthful description of the safe statistics of the real data\footnote{This could also be verified, assuming $\mathcal{G}$ produces datasets which match the safe statistics perfectly.}

Given the safe statistics $\psi$ of a real dataset, auditing a dataset card can be performed by evaluating $\generator$ at each point in the set of datasets whose safe statistics are $\psi$, and measuring whether the output of $\generator$ changes:
\begin{equation} \label{eq test}
   \text{test}_\text{all}(\mathcal{C}, \psi) = 1 \text{ iff } \Big[\generator(d) = \generator(d')~~\forall d,d' \text{ such that } \phi(d) = \phi(d') = \psi\Big]. 
\end{equation}
However, even if we could check every dataset, evaluating whether $\generator$ varies is challenging. Typically, for any training dataset $d$, we will not directly observe the distribution $\generator(d)\in \mathcal{P}(\datasetspace)$, but only samples from it. Furthermore, we are working with distributions over $\datasetspace$, which is typically very high-dimensional (for samples of fixed size $N$, the space of datasets is of dimension $O(N|\objectspace|)$), making this challenging numerically.

To simplify this problem, we make an additional assumption on the generator.
\begin{assumption}
$\generator$ corresponds to sampling records independently and identically from a distribution $\tilde\generator$ on $\objectspace$ which depends deterministically on the input dataset.
Formally, this means that the generator is defined by a map $\tilde \generator: \datasetspace \to \mathcal{P}(\objectspace)$ with (for a desired synthetic dataset of size $N$) the construction $\generator(d) = (\tilde \generator(d))^{\otimes N} \in \mathcal{P}(\objectspace^N)\subset \mathcal{P}(\datasetspace)$.
\end{assumption}
The key advantage of this assumption is that we can focus on the simpler map $\tilde\generator: \datasetspace \to \mathcal{P}(\objectspace)\equiv\datasetspace$. Rewriting in terms of our $\Theta$ statistics,  the auditing problem is equivalent to evaluating whether the restricted map
\begin{equation}\label{abstractregression}
\tilde\generator_\psi^\Theta:= \Theta\circ \tilde \generator\circ \Theta^{-1}:\underbrace{\Big(S_\Theta\cap \big(\{\psi\}\times \Phi^\perp\big)\Big)}_{\substack{\text{Statistics of valid datasets}\\\text{with safe statistics } \psi}}\to S_\Theta
\end{equation}
is constant (as defined in \eqref{eq test}), given (noisy) observations of its output. This reduces auditing a generator to a (nonlinear) regression problem of a map between spaces of statistics.

\subsection{Auditing a generator in practice}

The formal approach above raises two practical questions: ``How can we evaluate the  abstract map $\tilde\generator_\psi^\Theta = \Theta\circ \tilde \generator\circ \Theta^{-1}$?'', and ``How can we check if $\tilde\generator_\psi^\Theta$ is constant?''.

To evaluate $\tilde\generator_\psi^\Theta$ we will need to work with samples generated from $\tilde \generator(d)$, where we can choose the training datasets $d\in \datasetspace$. The formal approach above defines $\tilde\generator(d)$ as the distribution from which we sample observations, which is well approximated by the empirical distribution of a very large i.i.d.~sample. In practice, many of the generators we wish to consider will be randomised when training from data, so the samples we take will only be independent conditional on the training run. To account for this, we take a batched generation approach, where we will produce $K$ datasets, each of large size $n$, from independent training runs. Fixing a training dataset $d$ with safe statistics $\phi(d) = \psi$, our process is:
\begin{enumerate}[(i)]
    \item For each $k\leq K$, sample a dataset $D^{\mathrm{sym}}_k=\{x_1,..., x_n\}$ from the SDG $\tilde\generator(d)$, which we represent through its empirical  statistics $\Theta(D^{\mathrm{sym}})$. 
    \item Approximate\footnote{Higher moments of the distribution of $\{\Theta(D_k^\mathrm{sym})\}_{k\ge 0}$ may be of interest, when the generator produces only conditionally independent records. This corresponds to studying the full dataset generator $\generator$, rather than the individual record generator $\tilde \generator$. In practice, however, this may become computationally difficult, so we will focus on the average statistics.} the generating distribution $\tilde\generator_\psi^\Theta\big(\Theta (d)\big) \approx \frac{1}{K}\sum_k\Theta(D^{\mathrm{sym}}_k)$. If $\Theta$ depends on the empirical distribution linearly, this is exact in the limit $n,K\to\infty$.
\end{enumerate}
For notational convenience, for a vector $\beta$ of the same dimension as $\Theta(D)$, we define 
\begin{equation}\label{gdef}
g_\beta(D) := \beta^\top\Theta(D),
\end{equation}
so that, with $D^{\mathrm{sym}}$ as above, $\frac{1}{K}\sum_kg_\beta(D^{\mathrm{sym}}_k) \approx \beta^\top\tilde\generator_\psi^\Theta\big(\Theta (d)\big)$.

We now return to the problem of auditing a generator by considering the associated regression problem~\eqref{abstractregression}. We here propose a simple heuristic auditor that requires very few generator runs, but can effectively detect violation of generator cards. Practically, we will need to make some modelling assumptions on how  $\tilde \generator_\psi^\Theta$ depends on $\Phi^\perp$. These assumptions will suggest an appropriate test, however they are not critical to our conclusions: a generator which fails the audit test will still be known not to be decomposable, even if these modelling assumptions are false.

A natural first attempt to evaluate a generator is to model  $\tilde \generator_\psi^\Theta$ as linear in $\Phi^\perp$, and then use linear regression to search for the directions of greatest change. Choose an orthonormal\footnote{In practice, we may limit our attention to a (large) finite number of elements in this basis for computational purposes.} basis $(b_1, b_2, ..., b_{m-m_s})\subset \mathbb{R}^m$ of $\Phi^\perp$. Augmenting this with an orthonormal basis $(b_{m-m_s+1},..., b_m)$ of $\Phi$, we obtain a change of basis matrix $\mathcal{B}$, where $\mathcal{B}x = \sum_i x_i b_i$.
As per the dataset card, $\tilde \generator_\psi$ should not preserve the statistics in $\Phi^\perp$.
We thus model the generator as a map which \textit{preserves basis directions linearly}, that is, for a vector $a$ (which determines how different directions are scaled),
\[\tilde \generator_\psi^\Theta(x) = s_c + \mathcal{B}[\mathrm{diag}(a)] x.\]
Given this approximate model, our challenge is to estimate the vector $a$. However, as $\mathrm{diag}(a)$ is diagonal, it is enough to take a pair of datasets for which all statistics in $\Phi^\perp$ are varied, and then to estimate the coefficients $a$ through univariate linear regression in each coordinate. Our method uses two steps:

\textbf{Step 1:} We first seek to identify a direction in $\Phi^\perp$ in which the generator is varying. In order to get the best signal-to-noise ratio in our regression, we construct \textit{extremal datasets}: Begin with a  ``starting dataset'' $D^*$ with $\phi(D^*) = \psi$. For a randomly chosen direction $\beta \in \Phi^\perp$, we move $D^*$ as much as possible, to obtain $(D^-_\beta, D^+_\beta)$, such that (1) the unsafe statistics all change (i.e.~$b_i^\top\Theta(D^-_\beta) \neq b_i^\top\Theta(D^+_\beta) \text{ for all }i\leq m-m_s$) but (2) their safe statistics are identical $\phi(D^-_\beta) = \phi(D^+_\beta)=\psi$.
For a given dataset $d$, this is obtained by finding the largest $\alpha$ such that the resulting dataset is valid,
\begin{equation}
\label{eq:extremal_datasets}
D^{\pm}_\beta = \Theta^{-1}\big(\Theta(d) \pm \alpha^\pm_\beta \beta\big)~~\text{where } \alpha^{\pm}_\beta \in \mathop{\mathrm{arg\,max}}_\alpha \Big\{\alpha : \Theta(d) \pm \alpha \beta \in S_\Theta\Big\}.
\end{equation}
For each extremal dataset $D^\pm_\beta$, we sample large synthetic datasets $[D^\pm_{\beta}]^{\mathrm{sym}}_k$, for $k=1,...,K$, from trained generators $\generator(D^\pm_\beta)$.
Given our diagonal assumption, and using \eqref{gdef}, the regression estimate of the $i$th component of  $a$ is approximately (for $i\leq m-m_s$)
\begin{equation}
\label{eq:estimating_a}
    \hat{a}_i = \frac{1}{K}\sum_{k=1}^K\frac{g_{b_i}([D^+_{\beta}]_k^{\mathrm{sym}})- g_{b_i}([D^-_{\beta}]_k^{\mathrm{sym}})}{\alpha^+_\beta-\alpha^-_\beta}\approx \frac{\Big(\tilde\generator^\Theta_\psi(\Theta(D^+_\beta)) - \tilde\generator^\Theta_\psi(\Theta(D^-_\beta))\Big)^\top b_i}{\alpha^+_\beta-\alpha^-_\beta} .
\end{equation}
In principle, the auditor could make decisions based only on the estimates $\hat a$: if the coefficients of private statistics are significantly different from zero, then we know $\tilde \generator^\Theta_\psi$ is revealing private statistics.
However, this is unreliable due to randomness in the generation procedure.
Instead, we use these coefficients to select a unit vector in $\Phi^\perp$ along which $\tilde\generator^\Theta_\psi$ appears to be varying,
\begin{equation}
\label{eq:estimating-beta*}
    \beta^* = \frac{\sum_{i=1}^{m-m_s} \hat a_i b_i}{\sqrt{\sum_{i=1}^{m-m_s}\hat{a}_i^2}}.
\end{equation}

\textbf{Step 2:} We now evaluate whether the generator uses statistics collinear with $\beta^*$ when producing synthetic data.
For this, we generate extremal datasets in the direction $\beta^*$, $(D^-_{\beta^*}, D^+_{\beta^*})$, and train generators $\tilde\generator(D^\pm_{\beta^*})$, from which we sample large synthetic datasets $[D^\pm_{\beta^*}]^{\mathrm{sym}}_k$ for $k=1,..., K$.
Regressing $\tilde\generator^\Theta_\psi$ in the direction $\beta^*$ gives
\[\begin{split}\frac{\big(\tilde\generator^\Theta_\psi(\Theta(D^+_{\beta^*})) - \tilde\generator^\Theta_\psi(\Theta(D^-_{\beta^*}))\big)^\top \beta^*}{\alpha^+_{\beta^*}-\alpha^-_{\beta^*}} &\approx \frac{1}{\alpha^+_{\beta^*}-\alpha^-_{\beta^*}}\cdot\frac{1}{K}\sum_{k=1}^K\Big(g_{\beta^*}([D^+_{\beta^*}]^{\mathrm{sym}}_k) - g_{\beta^*}([D^-_{\beta^*}]^{\mathrm{sym}}_k)\Big).\end{split}\]
If the generator is constant, then this coefficient should be close to $0$.
Hence, we perform a two-sample test for mean equality
\begin{equation}\label{gtest}
\text{test}_\text{practical}(\mathcal{C}) = 1 \text{ iff } \bigg|\frac{1}{K}\sum_{k=1}^K g_{\beta^*}([D^+_{\beta^*}]^{\mathrm{sym}}_k) - \frac{1}{K}\sum_{k=1}^K g_{\beta^*}([D^-_{\beta^*}]^{\mathrm{sym}}_k)\bigg| > \tau.
\end{equation}
We perform this test as a two-sided $t$-test, allowing us to select a threshold $\tau$ to adjust the false positive/false negative error rates of the test and to get a measure of its statistical significance.

\begin{remark}
With this framework, we can see that the problem of auditing a generator can be seen as a hypothesis test. Our null hypothesis is that the generator is decomposable, and we build a test to see whether this hypothesis is false. If our test fails to detect additional leakage, this does not indicate that the generator is decomposable, but rather that we do not have evidence that it uses unsafe statistics. This distinction may be important in practical applications of privacy auditing, as it shows that a ``guarantee'' of decomposability is typically very difficult to obtain, but violations may be detectable.
\end{remark}

\subsubsection{Summary}

To summarise, our auditing procedure takes a generator card $\mathcal{C} = \left<\psi,\Phi,\generator\right>$ and assesses whether $\generator$ is decomposable in $\Phi$, i.e. whether $\generator$ only uses statistics $\psi = \phi(\realdata)$ to generate synthetic data.
The test operates in six steps:
\begin{enumerate}[(i)]
    \item Compute extremal datasets for a random direction $\beta$ (using the construction in \eqref{eq:extremal_datasets}).
    \item\label{firstK} Generate $K$ synthetic datasets from each extremal dataset.
    \item From these synthetic datasets, estimate the variation $\hat{a}_i$ in each direction in $\Phi^\perp$ using \eqref{eq:estimating_a}, and define a critical direction $\beta^*$ using \eqref{eq:estimating-beta*}.
    \item Compute extremal datasets for the direction $\beta^*$ (again using the construction in \eqref{eq:extremal_datasets}).
    \item Generate $K$ synthetic datasets from each extremal dataset (the $K$ here does not have to agree with the $K$ used in (\ref{firstK})).
    \item Calculate the two-sample test statistic \eqref{gtest} to assess whether the synthetic datasets differ in the direction $\beta^*$.
\end{enumerate}

\noindent Our construction makes the following assumptions on the generator:
\begin{enumerate}
    \item\label{assn:linear} The safe statistics form a linear space $\Phi$ (and hence $\Phi^\perp$ is also a linear space).
    \item\label{assn:iid} The generator computes a distribution over $\objectspace$, denoted $\tilde\generator$, from which synthetic records are sampled independently.
    \item\label{assn:det} The generator $\tilde\generator$ is a deterministic function of the training data.
    \item\label{assn:orthog} Unsafe statistics are replicated approximately linearly from the real data to the synthetic data.
\end{enumerate}

Assumption (\ref{assn:linear}) is key to our formalism, and is required to apply our framework.
Assumption (\ref{assn:iid}) is required for the auditing procedure to make sense. Assumption (\ref{assn:det}) is practically weakened by the use of a larger number of independent training datasets (larger value of $K$), and so is not critical if $K$ is large. Assumption (\ref{assn:orthog}) motivates our choice of test statistic, and is not critical to the conclusions drawn.

\section{Experiments on UK Census Data}
We now consider how the proposed framework can be applied in practice to inform the decisions of data controllers looking to share safe synthetic data. To motivate this, we provide examples of auditing applied to the Office for National Statistics' England and Wales 2011 Census Microdata Teaching File (the \textit{ONS Teaching File}) (\cite{ONSteaching}), an open access dataset containing an anonymised random sample of 1\% of responses from the 2011 Census of England and Wales, resulting in $\sim$570,000 records with 15 categorical attributes.
While our examples are only illustrative, and this sample data is already recognised as non-disclosive~(\cite{ONSteaching}), they demonstrate the breadth of applications under which our framework could be used as a part of disclosure control decisions.
We will consider three example analyses that can be carried out using this dataset, and for each example define a set of ``safe'' statistics $\Phi$.
We then evaluate how a range of synthetic generators (some decomposable in $\Phi$, and some not) perform in terms of utility, and apply the auditing framework to verify the decomposability of generators.

\subsection{Example Analyses}
\label{sec:use_cases}

We begin by specifying some notation. We write $\mathcal{A} = (A_1, \dots, A_d)$ for the list of attributes of the ONS Teaching File; since each attribute is categorical, they have associated finite domains $\objectspace_i = \{1, \dots, n_i\}$. We thus have $\objectspace = \objectspace_1 \times \dots \times \objectspace_d$.
We focus our attention on subsets of attributes $\mathcal{A}' \subset \mathcal{A}$, and the resulting $|\mathcal{A}'|$-way marginal distributions. Writing $x_{\mathcal{A}'}$ for the corresponding entries of an observation $x$, we parameterize the marginal distribution by the normalized counts of each combination of entries in $\mathcal{A}'$,
\[
\Theta_{\mathcal{A}'}(D) = \Big\{\frac{1}{|D|}\sum_{x\in D}I\{x_{\mathcal{A}'} = v\}\text{ for }v\in\objectspace_{a_1} \times \dots \times \objectspace_{a_l}\Big\}.
\]
Furthermore, define $\Theta^{(k)}_{\mathcal{A}'} = \{\Theta_{A} \text{ for } A \subset \mathcal{A}', |A| = k\}$ as the space of all $k$-way marginals from (subsets of) the set of attributes $\mathcal{A}'$.
We suppose that the analyst is interested in a specific subset of the attributes, but that a data controller is only willing to release some lower-dimensional marginal distributions.

\begin{application}[Summary Statistics] Adapting an example from the ONS Teaching File documentation~(\cite{teaching_file_possibilities}), we consider tabulated counts of economic activity, by occupation and gender. The analyst's aim is to examine the ``gender gap" (difference of proportion between sexes) for economically active people for each occupation. We suppose only 2-way tables can be released. In this setting we have variables of interest $\mathcal{A}_1^*$ and safe statistics $\Phi$, where
\begin{align*}
    \mathcal{A}^*_1 & = \left\{\texttt{Economic Activity}, \texttt{ Occupation}, \texttt{ Sex}\right\}, \quad
    \Phi = \linspan{\Theta^{(2)}_{\mathcal{A}^*_1}(d)}_{d\in \datasetspace}.
\end{align*}
\end{application}

\begin{application}[Inferential Modelling]
We consider a multinomial regression of Approximated Social Grade on Ethnic Group, Country of Birth and Family Composition.
As ``safe'' statistic $\Phi$, we take the 3-way marginal of the demographic attributes, and the pairwise marginals of each demographic attribute with the target attribute.
These statistics are such that they are sufficient for this analysis: the synthetic dataset should lead to the same statistical model as the real dataset.
\begin{align*}
    \mathcal{A}^*_2 &= \mathcal{A}^*_{2x}\cup\mathcal{A}^*_{2y}\\
    \mathcal{A}^*_{2x} &= \left\{\texttt{Ethnic Group}, \texttt{ Country of Birth}, \texttt{ Family Composition}\right\}\\
    \mathcal{A}^*_{2y} &= \left\{\texttt{Approximated Social Grade}\right\}\\
    \Phi &= \linspan{\Theta^{(3)}_{\mathcal{A}^*_{2x}}, \Theta^{(2)}_{\{a\}\cup\mathcal{A}^*_{2y}}~\text{ for } a \in \mathcal{A}^*_{2x}}
\end{align*}
\end{application}

\begin{application}[Machine Learning]
We consider a setting where more complex modelling approaches are applied. Specifically, we aim to build a random forest classifier to predict individuals' health status based on their socio-demographic attributes. As there are many variables, only pairwise (2-way) marginal distributions are considered ``safe''.
\begin{align*}
    \mathcal{A}^*_3 & = \Big(\texttt{Health}, \texttt{ Hours worked per week}, \texttt{ Age}, \texttt{ Marital Status},\\
    &\qquad \texttt{ Occupation}, \texttt{ Industry}, \texttt{ Region}\Big), \\
    \Phi & = \linspan{\Theta^{(2)}_{\mathcal{A}^*_3}(d)}_{d\in \datasetspace}.
\end{align*}
\end{application}

\subsection{Generators}
Several generators from prior work fall within the decomposability framework (see Section~\ref{sec:framework}).
We implement the following models, which allow an analyst to specify $\Phi$ as a combination of any number of $k-$way marginals:
\begin{enumerate}
    \item \textbf{Iterative Proportional Fitting (IPF)}, a model which estimates the data distribution by iteratively projecting a representation $T \in [0,1]^\objectspace$ over each known marginal, one at a time, until convergence. With univariate marginals, IPF is guaranteed to converge to a distribution of maximal entropy satisfying the constraints given by the marginals~(\cite{Cuturi2013}). This is a common tool in the literature on optimal transportation, and is has been used in some areas of synthetic data generation~\cite{Jeong2016}. 

    \item \textbf{Private-PGM}, a method which approximates the record distribution $T$ by a graphical model~(\cite{mckenna2021winning}). Private-PGM is designed to scale to much larger domains than IPF, and to be robust to noise addition, e.g.~to satisfy differential privacy (DP).
\end{enumerate}

We also compare these methods with generators that are not decomposable for a prespecified set of statistics $\Phi$.
\begin{enumerate}
    \item \textbf{MST}, a method based on Private-PGM which includes an additional step to automatically select a set of marginals to measure~(\cite{mckenna2021winning}). This method is designed to satisfy $(\varepsilon,\delta)-$DP for arbitrary $\varepsilon,\delta$.\footnote{Importantly, the first step of this algorithm computes a set of statistics to be measured. However, this set is computed using the data, and can vary for each run of the algorithm. Hence, MST does not satisfy the decomposability framework.}
    \item \textbf{PrivBayes}, which selects and trains a Bayesian network representation of the data distribution with $\varepsilon-$DP~(\cite{zhang2017privbayes}).
    \item \textbf{CTGAN}, a generative adversarial network (GAN) for tabular data~(\cite{xu2019modeling}). This does not provide theoretical privacy guarantees.
    \item \textbf{PATEGAN}, a GAN trained with $\varepsilon-$DP using the PATE framework~(\cite{jordon2018pate}).
    \item \textbf{SynthPop}, a model where attributes are generated sequentially, each attribute is generated conditional on all attributes generated before it using a statistical model that approximates this conditional distribution~(\cite{nowok2016synthpop}).
\end{enumerate}
For all methods satisfying differential privacy, we select high values for the privacy parameters ($\varepsilon=1000$ and $\delta=10^{-5}$) for fair comparison with methods not providing such guarantees. In order to avoid the curse of dimensionality (which is needed for IPF), for decomposable generators we produce synthetic data only for the subsets of attributes of interest for each application. For all general purpose methods, we generate synthetic datasets using all attributes, as these are assumed to be application-agnostic.

\subsection{Utility Evaluation}

For Application 1, we evaluate utility by reporting the RMSE between the 3-way marginal for attributes $\mathcal{A}^*_1$ of the ONS Teaching File, and of synthetic datasets obtained with each generator. We also calculate the RMSE for the gender gap, defined as the difference between the proportion of sexes in each occupation for economically active persons. The results of these tests are presented in Table~\ref{tab:results-use-case-1}.
We find that, with the exception of \texttt{SynthPop}, decomposable methods outperform the other methods, capturing the 3-way marginal with higher accuracy.

\begin{table}
\parbox{\textwidth}{%
    \caption{   \label{tab:results-use-case-1}
Utility analysis for Application 1: RMSE between the 3-way marginal of the real data and the synthetic data for the attributes $\mathcal{A}^*_1$ (center column), and RMSE between the gender gap disaggregated by occupation for economically active people.}
    \vspace{0.1cm}
\centering
\mbox{%
    \begin{tabular}{l|c|c}
         & \textbf{3-way Marginal} & \textbf{Gender Gap}\\
        \textbf{Generator} & \textbf{(RMSE)} & \textbf{(RMSE)}\\\hline
        IPF & $5.39\cdot 10^{-4}$ & $4.77 \cdot 10^{-2}$ \\
        Private-PGM & $5.35\cdot 10^{-4}$ & $4.94 \cdot 10^{-2}$\ \\ \hline
        MST$_{\varepsilon=1000}$ & $1.04 \cdot 10^{-3}$ & $8.68 \cdot 10^{-2}$\\
        PATEGAN$_{\varepsilon=1000}$ & $1.29 \cdot 10^{-2}$ & $0.570$ \\
        PrivBayes$_{\varepsilon=1000}$ & $8.97 \cdot 10^{-3}$ & $0.284$ \\
        CTGAN & $3.62 \cdot 10^{-3}$ & $0.170$\\
        SynthPop & $9.60 \cdot 10^{-5}$ & $4.61 \cdot 10^{-3}$\\
    \end{tabular}}}
\end{table}

\begin{table}
    \parbox{\textwidth}{%
    \caption{ \label{tab:results-use-case-2}
Utility analysis for Application 2: Accuracy of logistic regression trained on synthetic data, and evaluated on a holdout of the real and synthetic data. The first column measures the accuracy on the real data (how well a model trained on synthetic data transfers to real data), and the second is evaluated on synthetic data (how an an analyst would believe a model trained on synthetic data would perform).}
    \vspace{0.1cm}
\centering
\mbox{%
    \begin{tabular}{c|c|c}
         & \textbf{Accuracy on} & \textbf{Accuracy Estimated} \\
        \textbf{Generator} & \textbf{Real Data} & \textbf{on Synthetic Data}\\\hline
        Real Data & 0.341 & 0.341\\ \hline
        IPF & 0.341 & 0.341 \\
        Private-PGM & 0.341 & 0.340 \\ \hline
        MST$_{\varepsilon=1000}$ & 0.312 & 0.297 \\
        PATEGAN$_{\varepsilon=1000}$ & 0.205 & 0.403 \\
        PrivBayes$_{\varepsilon=1000}$ & 0.312 & 0.306 \\
        CTGAN & 0.288 & 0.380\\
        SynthPop & 0.341 & 0.340\\
    \end{tabular}}}
\end{table}

\begin{table}
    \parbox{\textwidth}{\caption{ \label{tab:results-use-case-3}Utility analysis for Application 3: accuracy of a random forest trained on a train split of the data, on (1) a test split of the real data, and (2) a test split of the synthetic data. We also report the Jaro similarity for the ordering of variable importance between the model trained on real data and that on synthetic data. This is estimated using permutation importance on the test split of real data.}
        \vspace{0.1cm}
    \centering
\mbox{%
    \begin{tabular}{c|c|c|c}
        & \textbf{Accuracy on} & \textbf{Accuracy Estimated} & \textbf{Similarity in} \\
        \textbf{Generator} & \textbf{Real Data} & \textbf{on Synthetic Data} & \textbf{Feature Ordering}\\\hline
        Real Data & 0.441 & 0.441 & 1.0 \\ \hline
        IPF & 0.434 & 0.436 & 0.98 \\
        Private-PGM & 0.438 & 0.370 & 1.0 \\ \hline
        MST ($\varepsilon=1000$) & 0.258 & 0.266 & 0.83 \\
        CTGAN & 0.329 & 0.383 & 0.89\\
        PATEGAN ($\varepsilon=1000$) & 0.210 & 0.460 & 0.49\\
        PrivBayes & 0.215 & 0.222 & 0.70 \\
        SynthPop & 0.437 & 0.473 & 0.98 \\
    \end{tabular}}}
\end{table}

For Application 2, we evaluate utility as the classification accuracy of the multinomial regression.
The statistical model is trained on a fraction ($\nicefrac{3}{4}$) of the synthetic dataset, and its accuracy is estimated on a holdout fraction ($\nicefrac{1}{4}$) of the ONS Teaching file, as well as on the remainder of the synthetic data $\synthdata$.
The former tells us how well a model trained on synthetic data performs on real data, while the latter describes the belief that an analyst might have in the accuracy of their model when given access to only the synthetic data. Ideally, we would like to see a similar levels of accuracy in the model applied to the synthetic data and observed data sets respectively. The results are presented in Table~\ref{tab:results-use-case-2}.

Here, we find that models trained on synthetic data from decomposable methods perform similarly to those trained on real data.
This is due to our choice of safe statistics: we have chosen $\Phi$ explicitly to contain all information needed to train a logistic regression.
By comparison, other methods tend to perform less well. In particular, we observe that GAN-based methods tend to lead to overconfident models -- models with significantly higher accuracy estimated on synthetic data than on real data.

For Application 3, similarly to Application 2, we measure the accuracy of the random forest on a holdout subset of the real data, as well as a subset of the synthetic data. We also estimate the importance of each feature using permutation importance~(\cite{breiman2001random}), and measure the Jaro distance between the ranking of features in a model trained on synthetic data and that of a model trained on real data. Feature importance is a useful tool for data analysts who want to understand how features relate to the target variable.
The baseline distance (average distance between two randomly shuffled strings of length 6) here is $0.70$. Results for this evaluation are presented in Table~\ref{tab:results-use-case-3}.

We find that, despite only having access to 2-way marginal distributions, decomposable generators outperform other generators, which struggle to capture the complex interactions exploited by random forests. In particular, decomposable models preserve the ordering of feature importance, for which PATEGAN and PrivBayes perform no better than random.

An important point to temper these results is that the decomposable models (IPF and Private-PGM) were trained for specific tasks involving relatively few attributes, whereas non-decomposable models were trained over the whole dataset.
As such, the latter models certainly give better utility across a wider range of tasks. In practice, this suggests that synthetic data should be curated for a a specific analytical task, and limitations of applying this data to wider tasks should be clearly communicated.

\subsection{Auditing Generators}

We now apply our auditing procedure to synthetic data generators for each of our three applications detailed above.
For this, we consider the two generators that fit within our decomposability framework (IPF and Private-PGM), and instantiate them either with only ``safe'' statistics $\Phi$ (so-called \textit{Honest} generators) or with $\Phi$ and some additional ``unsafe'' statistics $\Phi^+ \subseteq \Phi^\perp$ (so-called \textit{Dishonest} generators).
The specific $\Phi^+$ we add depends on the use case: for Application 1, $\Phi^+ = \linspan{\Theta^{(3)}_{\mathcal{A}^*_1}}$ (the full three way marginal); for Application 2,  $\Phi^+ = \linspan{\Theta^{(4)}_{\mathcal{A}^*_2}}$ (the full 4-way marginal); for Application 3, we take an arbitrary 3-way marginal, 
$\Phi^+ = \linspan{\Theta^{(3)}_{\{\texttt{Marital Status}, \texttt{ Occupation}, \texttt{ Hours worked per week}\}}}$.
We also consider the \texttt{SynthPop} generator, which has similar utility to IPF/MST, but which we do not expect to be decomposable for $\Phi$.

For each application and generator, we apply the test described in Eq.~\eqref{gtest} with $K=10$ 10 repetitions of the generator.
In Applications 1 and 2, the dimensionality of $\Phi^\perp$ is small enough (respectively, 81, and 448) that we can explicitly compute an orthnormal basis $b$.
In Application 3, $\dim{\Phi^\perp} \approx 1.6 \cdot 10^6$, and we instead sample a random subspace of $\Phi^\perp$, specifically generated by a subset of $\Theta^{(3)}_{\mathcal{A}_3^*}$ of size $1000$ taken uniformly at random.

\begin{figure}[ht]
    \centering
    \includegraphics[width=\textwidth]{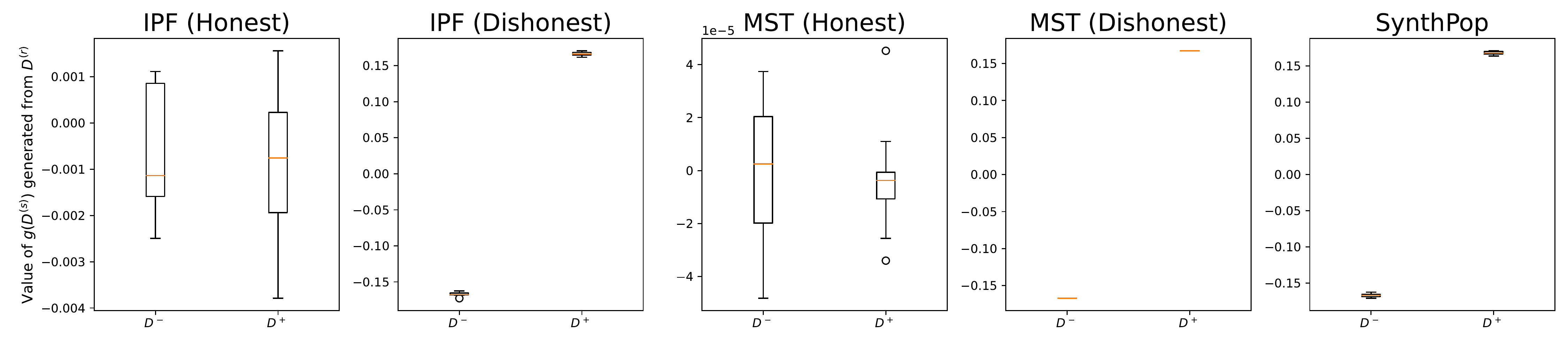}
    \includegraphics[width=\textwidth]{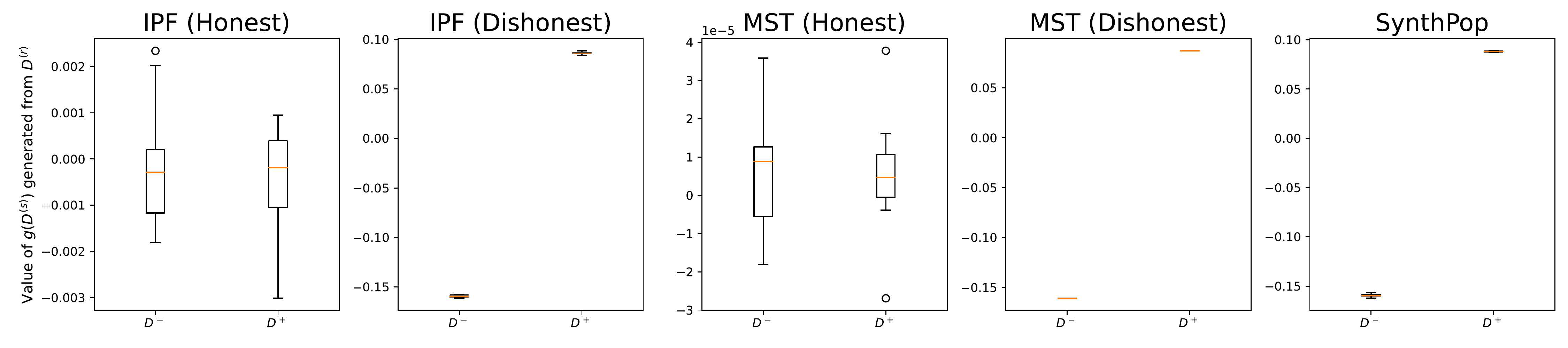}
    \includegraphics[width=\textwidth]{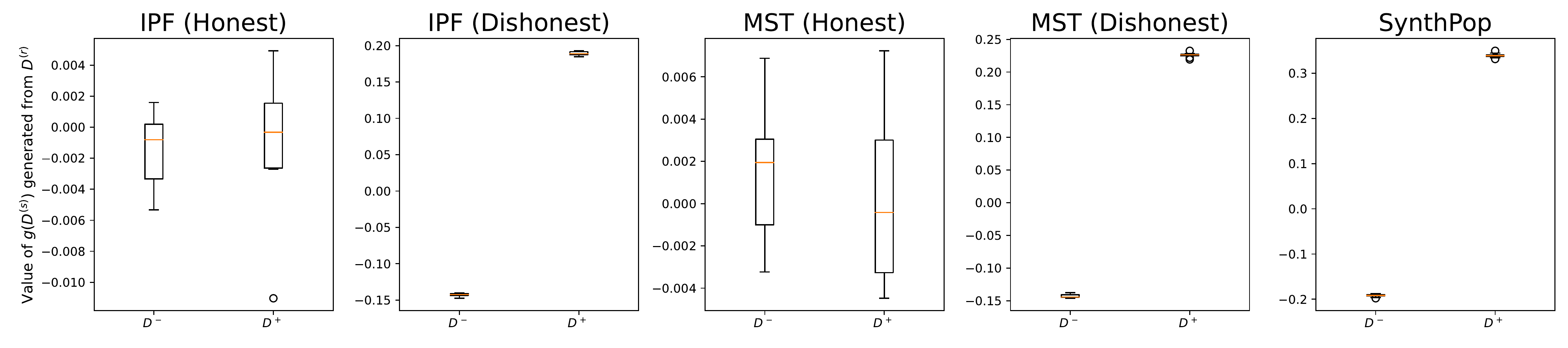}
    \caption{For each Application and generator, the distributions of the test statistic observations $g(D^\pm)$, as given in \eqref{gtest}.}
    \label{fig:auditing-use-case-1}
\end{figure}

In Fig.~\ref{fig:auditing-use-case-1}, we show the distribution of the samples for each application and generator. We observe that the difference in distribution is significant for all dishonest generators, and negligible for all honest generators.
We further perform a two-sided t-test to test whether the means of the distributions are identical. In Table~\ref{tab:p_value_auditing}, we show the p-values of this test. The results suggest that the auditing procedure effectively evaluates whether a dataset card is correct.
\begin{table}
\parbox{\textwidth}{%
    \caption{    \label{tab:p_value_auditing}
    p-values of a t-test for auditing, a lower value indicating that the mean value of $g([D^+_{\beta^*}]^{\mathrm{sym}}_k)$ differs from $g([D^-_{\beta^*}]^{\mathrm{sym}}_k)$, as given in \eqref{gtest}. The statistical test successfully identifies whether a generator card is correct (i.e.~the generator is honest) with high confidence.}
    \vspace{0.1cm}
    \centering
    \mbox{%
    \begin{tabular}{l|c|c|c}
        \textbf{Generator} & p-value (App.~1) & p-value (App.~2) & p-value (App.~3)\\ \hline
        IPF (Honest) & $0.54$ & $0.67$ & $0.67$\\
        IPF (Dishonest) & $8.9 \cdot 10^{-34}$ & $2.9 \cdot 10^{-37}$ & $5.7 \cdot 10^{-35}$ \\
        MST (Honest) & $0.89$ & $0.95$ & $0.47$ \\
        MST (Dishonest) & $5.4 \cdot 10^{-66}$ & $3.0 \cdot 10^{-69}$ & $2.3 \cdot 10^{-33}$ \\
        Synthpop & $5.1 \cdot 10^{-35}$ & $3.9 \cdot 10^{-38}$ & $3.7 \cdot 10^{-34}$      
    \end{tabular}
    }}
\end{table}

\section{Discussion}

The balance between utility and privacy for synthetic data is, in general, not well-understood.
In this paper, we argue that rather than attempting to develop high-fidelity private data, one should focus on utility: developing synthetic data for a limited set of tasks, and for which privacy concerns can be better focused.
We build on prior work to propose a framework where synthetic data is generated that matches a small set of well-chosen statistics. Datasets generated within this framework carry a ``generator card'', which transparently communicates the information used to generate it.

We then present a novel auditing methodology which enables data holders to verify whether the generator card is accurate, i.e.~whether the synthetic data only uses ``safe'' statistics. By incorporating this auditing framework within existing pipelines for generating synthetic data (such as that described in Section \ref{sec:framework}), we are able to build trust in the generation of synthetic data, both for data holders and analysts. 
The theory underpinning the framework allows it to be applied to any generator with an associated card. We show that, in a generic setting, the auditing procedure reduces to a hypothesis test in a regression problem, allowing us to adopt well-known approaches when determining whether a generator card is an accurate description. An interesting research direction would be to develop alternative statistical tests for comparing distributions over time series or networks that demonstrate the robustness of our general auditing framework.

\subsection*{Acknowledgements}
{\footnotesize
SC and LS acknowledge the support of the UKRI Prosperity Partnership Scheme (FAIR) under EPSRC Grant EP/V056883/1,
 and the Alan Turing Institute and the Office for National Statistics (ONS). RM acknowledges the support of the Office for National Statistics (ONS).
SC acknowledges the support of the Oxford--Man Institute for Quantitative Finance. 
}

\subsection*{Contributions}
{\footnotesize
FH, SC and LS proposed, formulated and formally analysed the framework and auditing methodology. ML, HW, OD, and RM proposed applications. FH and CM implemented the software. FH ran experiments and produced visualisations. FH, SC, LS, OD, RM, CM wrote the original draft. LS, SC and OD supervised the project.
}

%% Bibliography
\bibliographystyle{rss}
\bibliography{bibliography}

\end{document}